# UV imprinting and aligned ink-jet printing for multi-layer patterning of electro-optic polymer modulators

Xiaohui Lin,[1,*] Tao Ling,[2,*] Harish Subbaraman,[3] Xingyu Zhang,[1] Kwangsub Byun,[1] L. Jay Guo,[2] and Ray T Chen[1,**]

[1]Department of Electrical and Computer Engineering, The University of Texas at Austin, 10100 Burnet Rd, Austin, TX 78758
[2]Department of Electrical Engineering and Computer Science, University of Michigan, Ann Arbor, Michigan 48109
[3]Omega Optics, Inc., 10306 Sausalito Dr, Austin, TX 78759
* Equal contribution, **Corresponding author: raychen@uts.cc.utexas.edu

The present work demonstrates an electro-optic polymer based Mach-Zehnder (MZ) modulator fabricated utilizing advanced ultraviolet (UV) imprinting and aligned ink-jet printing technologies for patterning and layer deposition. The bottom electrode layer is designed and directly ink-jet printed on the substrate to form the patterned layer. The waveguide structure is formed into a bottom cladding polymer using a transparent flexible mold based UV imprinting method. All other layers can be ink-jet printed. The top electrode is aligned and printed over the Mach-Zehnder arm. The modulator demonstrates a V-pi of 8V at 3kHz. This technology shows great potential in minimizing the fabrication complexity and roll-to-roll compatibility for manufacturing low cost, light-weight, and conformal modulators at high throughput.

An Electro-Optic (EO) modulator is capable of encoding electronic signal on an optical carrier. The simplest kind of EO modulator uses a crystal such as Lithium Niobate [1], whose refractive index changes with an applied electric field. Besides, high performance modulator with 2.5ns switching time based on CMOS-compatible technology of hydrogenated amorphous silicon (a-Si:H) has been demonstrated[2]. Compared to above mentioned EO modulators, EO polymer based modulators are capable of even faster EO response[3] and negligible velocity mismatch between the optical and RF waves due to non-dispersive dielectric constant [4].

Traditionally, the fabrication process of the EO polymer modulator is complicated, involving several steps [5]. Multilayer patterning requires several lithography, deposition and lift-off processes, which increase the risk of device failure and lower the overall yield. Besides, typical channel waveguide or rib waveguide structures are fabricated by photolithography or electron beam lithography followed by reactive ion etching process. The disadvantages of this method include high surface roughness[6], low throughput[5] and high cost. Alternatively, imprinting technique can effectively overcome these shortcomings and provide the potential of roll-to-roll patterning capabilities at both micro- and nano-scales. Several optical components and devices have been demonstrated using various imprinting techniques, including EO polymer modulator [7], ultrasound detectors [8], micro-lens arrays [9], micro-ring resonators [10], LEDs [11], polarizers [12], etc. However, despite utilizing imprinting to pattern a layer, other material layers were still processed via spin coating, evaporation, etching, or lift-off methods, which will still increase the process complexity and cost. Therefore, we worked towards achieving an all-printable EO modulator. In the present work, in addition to a UV based imprinting method, which defines the waveguide channel with great repeatability, an ink-jet printing method is also integrated in the process to deposit the ground and the top driving electrodes. Note that, we have also tested the ink-jet printability of other layers, although standard spin coating process is used in this demonstration to simplify feasibility study. Therefore, our roll-to-roll compatible printing processes will simplify the fabrication process and enable high speed, low cost device fabrication. Especially from material consumption point of view, inkjet printing only deposits material at the desired region, thus minimizing both material consumption and wastage, compared with spin coating method. Also, it is featured by aligned printing, which reduces the complicated steps of lithography and lift-off. Furthermore, from system integration point of view, this method gives the flexibility of integrating several materials or printed components on a single platform, without increasing process complexities.

The printer employed in this research is a Fujifilm Dimatix Materials Printer (DMP-2800). It utilizes a piezoelectric cartridge to jet material onto a desired area on the substrate. With the aid of a fiducial camera, the printing region can be precisely defined within a positional error of 25µm.

A schematic of the Mach-Zehnder modulator we designed is shown in Fig.1. In our structure, we use an inverted rib waveguide structure to form the core. In the MZ modulator configuration, when a voltage is applied on one of modulating arms, the refractive index change in the EO polymer in that arm is governed by the equation $\Delta n = (-1/2)\gamma_{33} n^3 (V/d)$ and the phase difference between two arms is given by $\phi = (2\pi/\lambda) \Delta n L$, where $\gamma_{33}$ is the EO coefficient of the EO polymer, $n$ is the refractive index of the EO polymer, $V$ is the voltage applied, $d$ is the separation between ground and top electrodes, and $L$ is the modulation length. For our structure, the EO polymer (AJ-CKL1) has $\gamma_{33}$=80pm/V and $n$=1.63. The electrode separation and arm length are designed to be $d$=8.3µm and $L$=7.1mm, respectively. The theoretical value of the half-wave switching voltage is calculated to be $V_\pi = (\lambda\, d)/(L\, \gamma_{33}\, n^3) = 5.23V$.

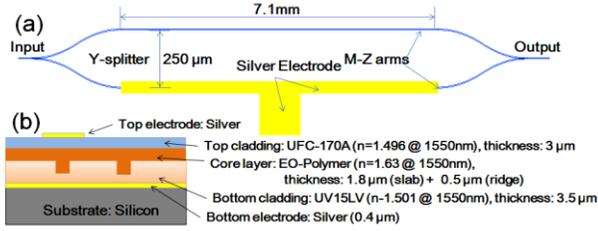

Fig.1 (a) Schematic top view showing the modulator structure, (b) Schematic cross section showing the different materials comprising the EO polymer modulator

In order to enable device development, the material system choice should meet certain criteria: 1) all the materials should satisfy the optical index requirements to form a waveguide, 2) the bottom cladding layer should be imprintable, 3) the core layer should have sufficient EO coefficient so that a suitable change in the index is achieved within a small applied E-field, and 4) the core material should have suitable viscosity to be ink-jet printed. In order to satisfy the physical and optical characteristics, we selected UV15LV (n=1.501@1.55μm) from MasterBond as the bottom cladding layer, EO polymer (AJ-CKL1, n=1.63@1.55μm) as the core layer and UFC-170A (n=1.496@1.55μm) from URAY Co. Ltd as top cladding layer. The choice of cladding materials is mainly based on the compatibility considerations of EO polymer. UV15LV is UV curable, solvent-free and also ink-jet printable. Thus it is a good candidate for roll-to-roll system. For the electrode layers, commercially available silver nanoparticle ink from Cabot Corp. was chosen.

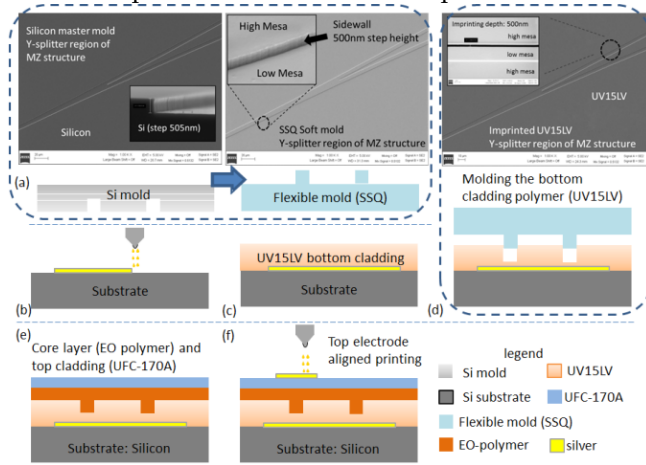

Fig.2 Main process flow for fabricating an electro-optic polymer modulator using imprinting and ink-jet printing method. (a) silicon and SSQ mold, (b) bottom electrode deposition, (c) bottom cladding layer deposition, (d) imprinting, (e) core layer and top cladding layer deposition, (f) top electrode deposition

The fabrication process is shown in Fig.2. First, a transparent epoxy Silsesquioxane (SSQ) mold is duplicated from silicon hard mold as shown in Fig.2(a). The feature step height is 0.5μm which equals the ridge height of the waveguide. The detailed mold fabrication process has been described before [13]. A 350~400nm bottom silver electrode layer along with alignment marks is ink-jet printed onto the silicon substrate, as shown in Fig.2(b). Then, UV15LV is deposited to form bottom cladding layer, as shown in Fig.2 (c). Next, the SSQ mold containing the MZ structure is brought into conformal contact with the bottom cladding layer, followed by UV imprinting and de-molding process, as shown in Fig.2(d). The EO polymer is then coated, followed by top cladding UFC-170A deposition, to form a rib waveguide structure, as shown in Fig.2(e). Finally, a top silver electrode is ink-jet printed on top of the top cladding layer, and aligned to one MZ arm as shown in Fig.2(f).

The UV imprinting process is used for transferring the pattern from the soft mold to the bottom cladding polymer (UV15LV). First, the fabricated transparent flexible SSQ mold is vapor-coated with anti-sticking layer and brought into conformal contact with a silicon wafer coated with a 4.18μm thick UV15LV layer. It is worth to mention that the transparent SSQ mold aids in aligning the waveguide pattern to the ink-jet printed bottom electrode pattern. After that, imprinting is performed in a pressure chamber with a pressure at 250 psi. Then, the imprinted UV15LV is fully cured by UV light for 5mins. Imprinting process produces inverse rib waveguide with dimensions of depth=500nm and width=4.75μm. The SEM image of a molded trench on UV15LV at the Y-splitter junction is shown in Fig.2(d).

After imprinting the MZ structure in the UV15LV bottom cladding layer, the EO-polymer is spin coated onto the bottom cladding layer to form a rib with 1.8μm thick in slab part and 0.5μm height in ridge part. After baking the EO-polymer, 3μm thick top cladding polymer is spin coated to finish the rib waveguide structure. Note that, although in the present work, the waveguide layers are deposited using spin-coating and patterned by imprinting, these layers can also be ink-jet printed with modified viscosity and surface tension. For example, for the UV15LV layer, we have previously shown its printability when demonstrating the thermo-optic switch [13].

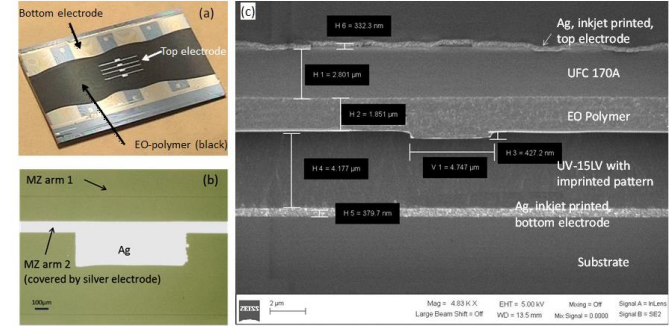

Fig.3 Microscope image of (a) printed EO polymer modulator, (b) ink-jet printed top electrode, scale bar: 100μm and (c) SEM picture of the modulator cross section, scale bar: 2μm

The key advantage in this fabrication technique is the utilization of ink-jet printing method for patterning both bottom/ground and top/driving electrode layers, without the need for evaporation or lift-off processes. With the aid of camera system and alignment mark located on both the ground electrode and the imprinted bottom cladding layers, we aligned and ink-jet printed the designed electrode pattern right on top of a modulating arm, with misalignment less than 10μm. Fig.3 (a) and (b) show the optical microscope pictures of a fabricated device and the printed top electrode, respectively. The length of the top electrode is 7.1mm. Fig.3 (c) shows a SEM cross-section image of all the layers that form the MZ modulator. The imprinted trench is filled with EO polymer, and the top and ground electrodes are uniformly deposited with the top electrode located right on top of the rib waveguide.

The electro-optic effect of EO polymer is created by contact poling process. Electric field of 80V/μm is applied across the silver electrodes. The temperature is controlled to rise from room temperature to peak poling temperature of 140°C, and then quickly decreased back to room temperature. The leakage current is monitored while poling.

In order to evaluate the modulation signal, the sample is mounted on an auto aligner for precise light coupling. TM-polarized light with 1.55μm wavelength from a tunable laser is launched into the input waveguide via a polarization maintaining lensed fiber, and the output light is collected by a single mode lensed fiber. The measured total insertion loss is around 26dB due to the high propagation loss of EO-polymer and the roughness of the cleaved input and output facets. Driving signal is applied across the driving and ground electrodes of the device, and the modulated optical signal is then collected by a photo detector connected to an oscilloscope. Fig.4(a) and (b) show the input and output when a 100kHz triangle wave signal is applied. Fig.4 (c) and (d) show the input and the modulator response, respectively, to a 15MHz sinusoidal signal. Due to the utilization of lumped electrode structure which is not specially designed for high speed purpose, the signal quality at higher speed was poor. By utilizing traveling wave electrodes, we expect to further increase the operating speed of the device [14]. The modulation depth at 10MHz is measured to be 10.3%, 12.5%, 20.4% and 25.5% when the applied bias voltages are 3V, 5V, 8V and 10V, respectively.

By fine tuning the voltage applied to the point of over modulation, the half-wave switching voltage $V_\pi$ is measured to be around 8.0V at 3KHz, which yields the $V_\pi \cdot L$=5.68V·cm. It is higher than the calculated value probably due to the low poling efficiency of the EO polymer, which reduces the in-device $\gamma_{33}$ value to *52.3pm/V*. Applying higher poling voltage may increase the risk of device break down. Although the presented $V_\pi \cdot L$ is higher than some of published values (e.g. 3.6V·cm[15], 1.8V·cm[3], 16.8V·cm[7], etc.), the device is still showing decent performance with less fabrication complexity. Furthermore, to the best of our knowledge, this is the first demonstration of a printed modulator.

In summary, we explicitly demonstrates a novel fabrication technology for polymer based EO modulator, taking advantage of UV imprinting to define waveguide patterns and ink-jet printing to deposit electrodes. It greatly reduces the fabrication complexity and lowers the cost. Modulation at 3kHz triangle signal is demonstrated, with $V_\pi$ of 8V. Since both the imprinting and ink-jet printing method are roll-to-roll compatible, our technology provides a practical way for fabricating low cost flexible optical devices with high throughput.

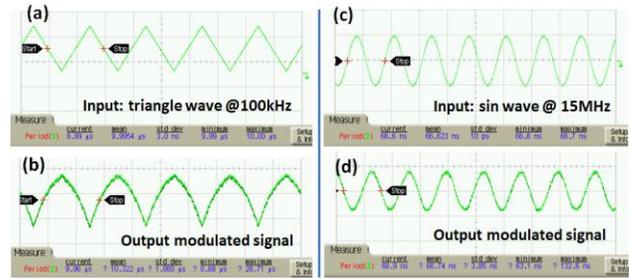

Fig.4 (a-b) 100kHz triangle wave input and corresponding modulator output signal. (c-d) 15MHz sinusoidal wave input and corresponding modulator output.

This work was supported by the Air Force Office of Scientific Research (AFOSR) Small Business Technology Transfer (STTR) contract FA9550-12-C-0052 monitored by Dr. Gernot Pomrenke. The fabrication were performed at the Microelectronics Research Center of The University of Texas at Austin and Lurie nanofabrication facility in University of Michigan, Ann Arbor.